\documentclass[twocolumn,showpacs,prd,floatfix,aps,nofootinbib]{revtex4}
\usepackage{latexsym}
\begin{document}
\title{Instantons and the infrared behavior of the fermion propagator in the Schwinger Model}
\author{Tomasz Rado\.zycki}
\affiliation{Department of Mathematical Methods in
Physics, Warsaw University, Ho\.za 74, 00-682 Warsaw,
Poland}
\email{torado@fuw.edu.pl} 

\begin{abstract}
Fermion propagator of the Schwinger Model is revisited from the point of view of its infrared behavior. The values of anomalous dimensions are found in arbitrary covariant gauge and in all contributing instanton sectors. In the case of a gauge invariant, but path dependent propagator, the exponential dependence, instead of power law one, is established for the special case when the path is a straight line. The leading behavior is almost identical in any sector, differing only by the slowly varying, algebraic prefactors. The other kind of the gauge invariant function, which is the amplitude of the dressed Dirac fermions, may be reduced, by the appropriate choice of the dressing, to the gauge variant one, if Landau gauge is imposed.
\end{abstract}
\maketitle

\section{Introduction}
\label{sec:intro}
Two dimensional and fully solvable field theoretical models have become efficient testing laboratories for various aspects of nonperturbative quantum field theory as well as in other branches of physics. One of the most productive examples of that kind  is the so called Schwinger Model (SM)~\cite{js} i.e. a system consisting of a massless fermion interacting with abelian gauge field in two space-time dimensions, together with its subsequent modifications (for instance $N$ fermionic flavors, fermion given a nonzero mass or chiral version of the model). Among its nontrivial and noteworthy features one can mention confinement, the existence of topological sectors, spontaneous chiral symmetry breaking and fermion condensate, gauge boson mass generation and axial anomaly. Beside these peculiarities, essential mainly to high energy physics, the SM, either massless or massive, of one flavor or more (as well as other models), has attracted the other researchers attention, for instance in the context of condensed matter physics~\cite{condens,tsvelik}. Here, much interest has also been payed to three dimensional models as $\text{QED}_3$. But $\text{QED}_2$ is of certain importance for the investigation of electrons in metals as well~\cite{tsvelik}.

There are plenty of papers dealing with different aspects of the SM, but relatively small concern~\cite{barci} has so far been devoted to the analysis of infrared anomalous dimensions of the fermionic correlation functions, which belongs to the both above areas of interest: confinement in quantum field theory and long-range correlations in condensed matter. The situation here is the more interesting, that all Green's functions, at least in the massless case, may be found explicitly thus allowing for the analytic and exact results~\cite{js,br,trjmn}. 

The question of field operators products in QFT having other than canonical dimensions, dates back to the Wilson's papers~\cite{wils}, where the author discovered this phenomenon in the Thirring model~\cite{thir} and $\phi^4$ theory. Since that time much work has been done to understand the behavior of the Green's or correlation functions in various systems, with special focus, at least in field theory, on the ultraviolet. The calculations have been mainly based on perturbation theory with further application of the renormalisation group methods. QED and QCD may serve here as important examples~\cite{anom}. In the present paper we would like to consider the SM fermion propagator (a fermion correlation function) and its anomalous dimension in the infrared. 

One of the known and important features of this model, making of it a valuable testing layout for certain aspects of QCD, is the existence of nontrivial topological sectors~\cite{cadam1,smil,gattr,maie,rot,gmc}. These topological sectors or instantons reveal their influence on expectation values of products involving fermion fields. This is connected with the fact, that objects of the kind $\Psi(x)\overline{\Psi}(y)$ are not chirally invariant and have nonzero matrix elements between distinct topological vacua~\cite{rajar}. Hence, if the true vacuum state (the so called $\theta$-vacuum) is defined through the sum, a Bloch-like superposition of topological vacua $|n>$ needed to maintain the cluster property~\cite{cdg}:
\begin{equation}
|\theta>=\sum_{n=-\infty}^{\infty}e^{in\theta}|n>\; ,
\label{eq:vacuum}
\end{equation}
the off-diagonal terms arising when one calculates the expectation values $<\theta|\hat{O}|\theta>$ not necessarily vanish.

The instanton background not only gives additional contributions to the propagator (or other fermion functions)~\cite{cadam1,trinst} but is likely to essentially change the long-range correlations between charged particles, due to the appearance of fermionic zero modes.. It is then worthwhile to examine the fermion propagator for large space-time or euclidean separations to find out, how values of anomalous dimensions depend on the topological sectors. Even if they are not of direct physical meaning because of their gauge dependence, the relative values between different sectors are interesting and their differences do not vary with the gauge parameter. As far as we know, no such calculation (which seems necessary for the complete infrared analysis of the propagator) has so far been performed. It is worth recalling here that for the four-point (purely fermion) function the essential modification of the bound state singularity in two-momentum space --- which is a simple pole as $k=0$ --- has been established in higher instanton sectors: $k=\pm 1,\pm 2$ (there appeared additional nonpolar --- branch point --- contributions)~\cite{trsing}. 

Although anomalous dimensions for fermion fields are gauge dependent quantities, they, as mentioned, eventually might be distinct in different topological sectors. Therefore, a natural question arises whether one is able to make all noncanonical exponents vanish, by the appropriate choice of gauge. Does a kind of Fried-Yennie gauge exist for all sectors or for each sector separately? The SM is a perfect tool to answer these questions since the fermion propagator is known in all sectors, thanks to the calculations performed in the euclidean space. To that aspect we devote section~\ref{sec:gvar}. 

Another interesting task is to find a gauge independent function and analyse its long-range behavior. Such functions are of certain importance in condensed matter physics but should also constitute essential objects in investigations concerning fundamental problems of QCD, like confinement for instance, similarly as Wilson loop does. It is well known, that analytical properties of the ordinary (gauge variant) function are gauge dependent. For example in QED, the mass-shell singularity of the electron propagator in $p$-space which, in one gauge, is a simple pole, in another becomes a branch point. 

Gauge invariant function, however, is not an unequivocal notion. In section~\ref{sec:path} we use the function deriving from Schwinger~\cite{jspath} (see the footnote in the quoted work) and~\cite{js} in the form:
\begin{equation}
\label{eq:sinvp}
<0|T(\Psi(x)\exp\left[-ig\int_{\gamma(x,y)}\!\!\!\!\!\!\!\!\!\! dz_\mu A^\mu(z)\right]\overline{\Psi}(y))|0>\; ,
\end{equation}
where ${\gamma(x,y)}$ is a certain space-time path from $x$ to $y$. The function~(\ref{eq:sinvp}) has been widely used in the literature for various systems~\cite{barci,mitra,ratwen,kh,gusy,ye}, sometimes with contradictory results. One may hope that this kind of an amplitude describes the propagation of physical particles. However, one has to keep in mind that (\ref{eq:sinvp}) is a path dependent object and, therefore, not unique. Considering the simplest case of a straight line, (probably more suitable for the ultraviolet asymptotics) which is tractable analytically, we find the change of the infrared behavior from power law decay to exponential one. This effect was shown to be existent (but cutoff dependent) and actually path insensitive in $\text{QED}_3$~\cite{mitra} and also demonstrated in the Schwinger Model~\cite{barci}, although without considering other than $k=0$ instanton sectors. We show, yet, that it appears independently of the gauge-field topology. A kind of an average over different paths (but still for $k=0$) was performed in~\cite{barci} showing a slightly faster decay, however without changing the leading exponential behavior.

In section~\ref{sec:dress} we consider another type --- although not entirely unrelated --- of the gauge independent function, defined as the vacuum amplitude of the dressed (and gauge invariant) fermion fields of the Dirac type~\cite{dir}: 
\begin{equation}
\label{eq:dres}
\tilde{\Psi}(x)=e^{-i\int d^2z{\cal F}_\mu(z,x)A^\mu(z)}\Psi(x)\; , 
\end{equation}
where ${\cal F}_\mu$ satisfies $\partial^\mu_z{\cal F}_\mu(z,x)=g\delta^{(2)}(z-x)$\footnote{Dirac's original suggestion for the static charge in the $3+1$ dimensional electrodynamics, was to take the dressing in the Lorentz noncovariant --- but local in time --- form: $$\tilde{\Psi}(t,\mathbf{x})=\exp\left[-\frac{ig}{4\pi}\int d^3y\frac{1}{|\mathbf{x} - \mathbf{y}|}\mathbf{\nabla}\mathbf{A}(\mathbf{y})\right]\Psi(t,\mathbf{x})\; .$$ Later this formula was modified to account for moving electrons~\cite{lav}}. Our choice of ${\cal F}_\mu$ reduces the propagator, again with all topological contributions taken into account, to that considered in section~\ref{sec:gvar}, provided the Landau gauge was chosen. This is due to the fact that the ${\cal F}_\mu$ function used, is purely  longitudinal and couples only to $\partial A$, which disappears in the Landau gauge~\cite{mitra}. As we will see in section~\ref{sec:dress}, the choice of the function ${\cal F}_\mu$ is related to the definition of the string in~(\ref{eq:sinvp}).

\section{Gauge variant propagator}
\label{sec:gvar} 
As written in the Introduction the original Schwinger Model consists of a massless fermion field $\Psi$ interacting with a gauge boson $A^{\mu}$. The Lagrangian density for this model has the following form:  
\begin{eqnarray} 
{\cal L}(x)&=&\overline{\Psi}(x)\left(i\gamma^{\mu}\partial_{\mu}
- g\gamma^{\mu}A_{\mu}(x)\right)\Psi (x)\nonumber\\* &-&
\frac{1}{4}F^{\mu\nu}(x)F_{\mu\nu}(x)\; , 
\label{eq:lagr} 
\end{eqnarray} 
with a gauge-fixing term, if needed.

The space-time has one time and one space dimension. It is then natural to choose the fermion field in the simplest form of the two-component spinor, this size being determined, for the even number of dimensions $D$, by the value $2^{D/2}$. In this case, for the $2\times 2$ Dirac gamma matrices one can choose the representation given below: 
\begin{equation} 
\gamma^0=\left(\begin{array}{lr}0 & \hspace*{2ex}1 \\ 1 & 0 
\end{array}\right)\; , \;\;\;\;\; 
\gamma^1=\left(\begin{array}{lr} 0 & -1 \\ 1 & 0 
\end{array}\right)\; ,
\label{eq:gamma}
\end{equation}
$$
\gamma^5=\gamma^0\gamma^1=\left( 
\begin{array}{lr} 1 & 0 \\ 0 & -1 \end{array}\right) \; .
$$ 
For the metric tensor we use the convention:
$$
g^{00}=-g^{11}=1\; ,
$$ 
and the antisymmetric symbol $\varepsilon^{\mu\nu}$ is defined by:
\begin{equation}
\varepsilon^{01}=-\varepsilon^{10}=1\; ,\;\;\;\;\; 
\varepsilon^{00}=\varepsilon^{11}=0\; ,
\label{eq:epsil} 
\end{equation}
but whenever it is required by the mathematical rigor, the considering of the corresponding euclidean space is assumed.

The nonperturbative formula for the fermion propagator was given already in the first Schwinger papers~\cite{js} and later rederived by other methods. It may be given the following form:
\begin{eqnarray}
&&S_{(0)}(x-y)=\label{eq:s0}\\
&&={\cal S}_0(x-y)e^{-\frac{i}{2}\int d^2ud^2w {\cal J}^\mu(u;x,y)\triangle_{\mu\nu}(u-w){\cal J}^\nu(w;x,y)}\; ,\nonumber
\end{eqnarray}
where $\triangle_{\mu\nu}(u-w)$ is a massive (of mass squared equal to $g^2/\pi$) vector boson propagator, in general gauge dependent, i.e. the propagator of the Schwinger boson. The current ${\cal J}^\mu$ may be written as:
\begin{eqnarray}
\label{eq:currj}
{\cal J}^\mu(z;x,y)&=&g\gamma^\mu\gamma^\nu\partial^z_\nu [D(z-x)- D(z-y)]=\\
&=&g(g^{\mu\nu}+ \gamma^5\epsilon^{\mu\nu})\partial^z_\nu[D(z-x)- D(z-y)]\; .\nonumber
\end{eqnarray}
It can be easily checked, with the use of: 
$$\Box_zD(z)=\delta^{(2)}(z)\; ,$$
that this auxiliary current is not conserved and describes the creation of a charged particle at $x$ and its anihilation at $y$:
\begin{equation}
\label{eq:divj}
\partial^z_\mu {\cal J}^\mu(z;x,y)=g(\delta^{(2)}(z-x)-\delta^{(2)}(z-y))\; .
\end{equation}
The symbol ${\cal S}_0(x)$ denotes the free fermion propagator, i.e. ${\cal S}_0(x)=-\frac{1}{2\pi}\frac{\not\!\!\;\, x}{x^2-i\varepsilon}$, and index $(0)$ on the l.h.s of (\ref{eq:s0}) refers to the trivial ($k=0$) instanton sector.

\subsection{Landau gauge}
\label{sec:gvar:subsec:lg}
If one chooses $\triangle_{\mu\nu}(x)$ to be transverse function, i.e. satisfying $\partial_x^\mu\triangle_{\mu\nu}(x)=0$, as Schwinger in his original papers did, one arrives to the following, well known, form of the propagator:
\begin{equation}  
S_{(0)}(x-y)={\cal S}_0(x-y)\exp\left[-ig^2\beta(x-y)\right]\; ,  
\label{eq:propbet} 
\end{equation} 
where we introduced a function $\beta$ defined by the two-momentum integral and expressible through cylindrical functions: 
\begin{widetext}
\begin{eqnarray}
\beta(x-y)&=&\int\frac{d^2p}{(2\pi)^2}\left(1-e^{ip(x-y)}\right)
\frac{1}{(p^2-g^2/\pi
+i\epsilon)(p^2+i\epsilon)}=\nonumber\\
&&\nonumber\\ 
&=&\left\{\begin{array}{ll}\frac{i}{2g^2}\left[-
\frac{i\pi}{2}+\gamma_E
+\ln\sqrt{ g^2(x-y)^2/4\pi}+ 
\frac{i\pi}{2}H_0^{(1)}(\sqrt{g^2(x-y)^2/\pi})\right] & \hspace*{3ex}
 x-y\;\;\;\; \text{timelike}\\ 
\frac{i}{2g^2}\left[\gamma_E+\ln\sqrt{-g^2(x-y)^2/4\pi}+K_0(\sqrt{-
g^2(x-y)^2/\pi})
\right]  & \hspace*{3ex}
 x-y\;\;\;\; \text{spacelike}\; .\end{array}\right.
\label{eq:beta}  
\end{eqnarray} 
\end{widetext}
Symbol $\gamma_E$ denotes here the Euler constant and functions
$H_0^{(1)}$ and $K_0$ are Hankel function of the first kind, and Basset
function respectively.

From~(\ref{eq:propbet}) and~(\ref{eq:beta}) one can easily deduce the IR asymptotics, considering euclidean or spacelike infinity. The canonical dimension is associated with the free field and, therefore, appears already in the coefficient ${\cal S}_0(x)$. The anomalous one comes from the infrared behavior of the function $\beta$. For $(x-y)^2\rightarrow -\infty$ this behavior is dictated by the logarithmic function, since $K_0$ is exponentially decreasing for large arguments. The leading term takes the form:
\begin{equation}
\label{eq:as0}
S_{(0)}(x-y)\sim \frac{e^{\gamma_E/2}}{\sqrt{2}}{\cal S}_0(x-y)(-g^2(x-y)^2/\pi)^{1/4}\; .
\end{equation}
In this case the anomalous dimension is equal to $-1/2$ and this reflects the known infrared behavior of $S_{(0)}(p)$ in momentum space: $\frac{\text{const}}{(-p^2)^{5/4}}$~\cite{stam,trpert}. The long-range correlation is then enhanced with respect to the noninteracting case, contrary to, what is called, Luttinger-type behavior. The obtained value, negative in the Landau gauge, may, however, be freely modified by a gauge transformation.

By the proper modification and extension of the formula (\ref{eq:s0}) (spoken of in detail in~\cite{trinst}) for the case with instantons, where zero modes of the Dirac operator contribute to the propagator (for the sector $k$, the number of these modes is equal to $|k|$~\cite{as}), one arrives to the formula for the $k=\pm 1$ sectors:
\begin{equation}
S_{(1)}(x-y)=\frac{ig}{4\pi^{3/2}}\left(\cos\theta-i\gamma^5\sin\theta\right)e^{\gamma_E 
+ig^2\beta(x-y)}\; .
\label{eq:s1}
\end{equation}
The angle $\theta$ comes from the definition of the true, physical vacuum of the theory given in~(\ref{eq:vacuum}). The formula~(\ref{eq:s1}) with the use of the infrared approximation of~(\ref{eq:beta}) leads to:
\begin{equation}
\label{eq:as1}
S_{(1)}(x-y)\sim \frac{ig e^{\gamma_E/2} e^{-i\gamma^5\theta}}{(2\pi)^{3/2}}\cdot \frac{1}{(-g^2(x-y)^2/\pi)^{1/4}}\; .
\end{equation}
Although in the spinor space the matrical structures of~(\ref{eq:as0}) and of~(\ref{eq:as1}) are different, their scaling properties are identical, so the anomalous dimension in this sector is still equal to $-1/2$. The asymptotic (noncanonical) scale invariance is then maintained. One should emphasize here, that apart from~(\ref{eq:as0}) and ~(\ref{eq:as1}), no other algebraic terms arise. The long-range expansion of (\ref{eq:beta}) shows, that subleading terms have an exponential character deriving from the behavior of the Basset function.

\subsection{Longitudinal contributions}
\label{sec:gvar:subsec:lc}
In this section we verify, how gauge transformations influence the obtained values of anomalous dimensions in all topological sectors. For the trivial sector $k=0$, the simplest idea is to move back to~(\ref{eq:s0}) and replace $\triangle_{\mu\nu}(x)$, which previously was chosen to be transverse, with the quantity:
\begin{equation}
\label{eq:dlong}
\triangle_{\mu\nu}(x)=\triangle^T_{\mu\nu}(x)\rightarrow \triangle^T_{\mu\nu}(x)+\triangle^L_{\mu\nu}(x)\; ,
\end{equation}
and see the dependence of the results upon different choices of $\triangle^L_{\mu\nu}(x)$. Using $\triangle^L_{\mu\nu}(x)=\partial_\mu\partial_\nu F(x)$, one can easily find from~(\ref{eq:s0}) and~(\ref{eq:divj}), that there would appear in~(\ref{eq:propbet}) an additional, multiplicative factor, which reads:
\begin{equation}
\label{eq:factor}
\exp[-ig^2(F(x-y)-F(0))]\; .
\end{equation}
Its form is dictated by the longitudinal part of the current ${\cal J}^\mu$. If we analyze the formula~(37) of~\cite{trinst} together with~(\ref{eq:divj}) and~(\ref{eq:divk}), it becomes obvious that~(\ref{eq:factor}) will be common for each instanton sector, since, while passing from $k=0$ to $k=\pm 1$, all terms additionally appearing in the exponent are transverse and, therefore, not sensible to the substitution~(\ref{eq:dlong}). 

The above result agrees with that obtained by Landau and Khalatnikov~\cite{lk} and later rederived by Zumino~\cite{zumino} for ordinary QED. The topological sectors are absent in $\text{QED}_{3+1}$ and, obviously, were not consiered at the time, but the transformation property described by~(\ref{eq:factor}) turns out to be general. 

Naturally, there exists a wide class of functions $F(x)$, one can select, but we restrict ourselves to those arising in $R_\xi$ gauges, with an unessential modification. We assume then:
\begin{equation}
\label{eq:f}
F(x-y)=\xi\int d^2w D(x-w)\triangle(w-y)\; ,
\end{equation}
where $\triangle(w-y)$ is a 2D Klein-Gordon propagator for a particle of certain mass, which, without a real change of the infrared asymptotics, may be chosen to be equal to the mass of the Schwinger boson: $g/\sqrt{\pi}$. With this choice, it can easily be verified that the additional factor~(\ref{eq:factor}) reduces to $\exp[i\xi g^2\beta(x-y)]$ leading to the asymptotics:
\begin{eqnarray}
S_{(0)}(x-y)\sim &&\left(\frac{e^{\gamma_E/2}}{\sqrt{2}}\right)^{1-\xi}{\cal S}_0(x-y)\nonumber\\
&&\times(-g^2(x-y)^2/\pi)^{(1-\xi)/4}\; ,\nonumber\\
\label{eq:asl}\\
S_{(1)}(x-y)\sim &&\left(\frac{e^{\gamma_E/2}}{\sqrt{2}}\right)^{1-\xi}\frac{ig e^{-i\gamma^5\theta}}{2\pi^{3/2}}\nonumber\\
&&\times \frac{1}{(-g^2(x-y)^2/\pi)^{(1+\xi)/4}}\; .\nonumber
\end{eqnarray}
As may be seen above, thanks to the universality of the factor~(\ref{eq:factor}), there exists a gauge $\xi=1$, for which in both sectors the anomalous dimensions disappear and the propagator recovers its canonical dimension. What is more, in this gauge $S_{(0)}(x-y)={\cal S}_0(x-y)$. This is easily understandable on the basis of perturbation calculation~\cite{trpert}. For $\xi=1$, the gauge boson propagator becomes proportional to $g_{\mu\nu}$. Therefore, if one draws any Feynman diagram with two external fermion `legs', each internal line containing this function and joining two vertices leads to the summation of the kind
\begin{equation}
\label{str}
\gamma^{\mu}\gamma^{\alpha_1}\gamma^{\alpha_2}\cdot ...  
\cdot \gamma^{\alpha_{2k+1}}\gamma_{\mu}\; ,
\end{equation}  
where $\gamma^{\alpha_1},\gamma^{\alpha_2}, ...  
\gamma^{\alpha_{2k+1}}$ come from either vertices or fermion propagators (remember that the free fermion propagator is proportional to $\gamma$) squeezed between these two vertices. The number of those $\gamma$'s is always odd (the number of fermion propagators always exceeds by one the number of vertices). Consequently, the corresponding expression vanishes, since any odd number of gamma matrices may, in two dimensions, always be
reduced to only one, by subsequent applications of the formula:
\begin{equation}
\gamma^\alpha\gamma^\beta\gamma^\rho=g^{\alpha\beta}\gamma^\rho+\epsilon^{\alpha\beta}\gamma^5\gamma^\rho=g^{\alpha\beta}\gamma^\rho+\epsilon^{\alpha\beta}\epsilon^{\rho\lambda}\gamma^\lambda\; .
\label{eq:gamred}
\end{equation}
After the reduction we simply apply $\gamma^{\mu}\gamma^{\alpha}\gamma_{\mu} =0$, which can also be deduced from~(\ref{eq:gamred}). 

On the other hand $S_{(1)}(x-y)$ remains nontrivial even in this gauge. Due to its nonperturbative and topological nature, this simple reasoning based of perturbative Feynman diagrams may not be applied.

\section{Gauge invariant, path dependent propagator}
\label{sec:path}
One of the candidates for a gauge invariant one fermion Green's function is that originating from the Schwinger's paper~\cite{jspath} (some kind of `string expression' appeared in the Dirac's paper too~\cite{dir}):
\begin{eqnarray}
\label{eq:sinvp1}
&&S^{inv}(x-y)=\\
&&=<0|T(\Psi(x)\exp\left[-ig\int_{\gamma(x,y)}\!\!\!\!\!\!\!\!\!\! dz_\mu A^\mu(z)\right]\overline{\Psi}(y))|0>\; .\nonumber
\end{eqnarray}
The gauge independence of $S_{inv}(x-y)$ my be easily verified, yet this quantity depends now on the chosen path $\gamma(x,y)$. Two questions come immediately into view. One is practical: for which path is it possible to analytically evaluate the vacuum expectation value~(\ref{eq:sinvp1})? The other is more fundamental: is there any physical justification for the choice of one $\gamma$ before another?

In practice the calculation of~(\ref{eq:sinvp1}) is possible only for the simplest choices, as straight line for instance. Such a path is certainly a good option for the ultraviolet asymptotics, since then $x$ is close to $y$ and the curve may be approximated with a short section of a straight line. For the infrared asymptotics we need infinitely long path, but, on the other hand, the long-range behavior should not depend too strongly on the contour peculiarities for finite values of $z$. This is to some extent supported by the results of~\cite{barci}.

The choice of the path $\gamma$, as well as of any other form of the invariant propagator, is related to the selection of the gauge for the noninvariant function. For instance the straight line path corresponds to the Fock gauge $(z_\mu-x_\mu)A^\mu(z)=0$~\cite{fock} since, in this case, the string in~(\ref{eq:sinvp1}) disappears and both invariant and noninvariant functions equal each other. This means that the gauge independence of~(\ref{eq:sinvp1}) becomes somewhat illusory. This is as if one told that the propagator calculated in certain fixed gauge is gauge independent. Obviously it is! Similarly as the position of the moving body in a fixed moment $t=t_0$ is time independent. Therefore the main point is to ascertain which path is the most convenient one (which might also mean that certain gauge or class of gauges is privileged). On the other hand, this choice can be ambiguous. It is likely that each type of experiment or measurement has its own preferable path and, therefore, one is unable to define it in a versatile way. Only a set of universal features for a unconfined particle is guaranteed to be preparation independent. One should not expect then, that all measurements performed over an electron (charged particle) obtained by the ionization of an atom with strong laser impulse give identical outcome as those over a trapped particle in the solid body. 

Nevertheless the propagator defined in~(\ref{eq:sinvp1}) has found its applications~(see for instance \cite{ratwen, ye, kanaya}) and, certainly, is worth being investigated with the topological contributions taken into account. Without going too deeply into above questions, we restrict ourselves, in the present work, to the segment of the straight line enclosed between $x$ and $y$, for which~(\ref{eq:sinvp1}) may formally be calculated in all instanton sectors.

It will be convenient to write down the string expression in the source form:
\begin{eqnarray}
\label{eq:str}
&&\exp\left[-ig\int_{\gamma(x,y)}\!\!\!\!\!\!\!\!\!\! dz_\mu A^\mu(z)\right]=\\
&&\;\;\;\;\;\;\; =\exp\left[-i\int d^2z{\cal C}_\mu(z;x,y)A^\mu(z)\right]\; ,\nonumber
\end{eqnarray}
where we introduced a certain new current:
\begin{equation}
\label{eq:cur}
{\cal C}_\mu(z;x,y)=g\int_{\gamma(x,y)}\!\!\!\!\!\!\!\!\!\! dw_\mu\delta^{(2)}(z-w)\; ,
\end{equation}
satisfying:
\begin{equation}
\label{eq:divc}
\partial^z_\mu {\cal C}^\mu(z;x,y)=g(\delta^{(2)}(z-x)-\delta^{(2)}(z-y))\; .
\end{equation}
It is worth noting that this divergence is identical to~(\ref{eq:divj}), which means that the difference of the two currents ${\cal J}^\nu$ and ${\cal C}^\mu$ is transverse:
\begin{eqnarray}
\label{eq:jmc}
&&{\cal J}^\mu(z;x,y)-{\cal C}^\mu(z;x,y)=\\
&&=g\left[\gamma^5\epsilon^{\mu\alpha}\partial^z_\alpha\partial_z^\beta-(\Box_zg^{\mu\beta}-\partial_z^\mu\partial_z^\beta)\right]
\int_{\gamma(x,y)}\!\!\!\!\!\!\!\!\!\!du_\beta D(z-u)\; .\nonumber
\end{eqnarray}
As we will see below, this transversality guarantees the gauge invariance of the final expression. Now we calculate the fermion propagator using functional integral formalism, where it is defined as the derivative over fermionic sources $\eta$ and $\overline{\eta}$ of the generating functional:
\begin{eqnarray} 
\label{eq:gener} 
&&Z(\eta,\overline{\eta},J)=\sum_{k=-
\infty}^{\infty}e^{ik\theta}Z^{(k)}[\eta ,\overline{\eta}, J]=\\
&&\;\;\;\;\; =\sum_{k=-
\infty}^{\infty}e^{ik\theta}\int D\Psi D\overline{\Psi} DA^{(k)} \exp\bigg[i\int d^2x 
({\cal L}^{(k)}(x) \nonumber\\
&&\;\;\;\;\;  +\; \overline{\eta}(x)\Psi (x) + 
\overline{\Psi}(x)\eta (x) + J^{\mu}(x)A^{(k)}_{\mu}(x))\bigg]\; , \nonumber
\end{eqnarray}
where $A^{(k)}_\mu$ is the gauge field restricted to the $k$-instanton sector, and ${\cal L}^{(k)}$ is the Schwinger Model Lagrangian density with the substitution $A_\mu\rightarrow A^{(k)}_\mu$. One immediately sees, that the only modification introduced by the string~(\ref{eq:str}) for $k=0$ is the replacement of the current $J_\mu$ with $J_\mu-{\cal C}_\mu$. This results in the following change of the formula~(\ref{eq:s0}):
\begin{widetext}
\begin{equation}
\label{eq:sinv0}
S^{inv}_{(0)}(x-y)={\cal S}_0(x-y)
\exp\left[-\frac{i}{2}\int d^2ud^2w ({\cal J}^\mu(u;x,y)-{\cal C}^\mu(u;x,y))
\triangle_{\mu\nu}(u-w)({\cal J}^\nu(w;x,y)-{\cal C}^\mu(w;x,y))\right]\; .
\end{equation}
The evaluating of the exponent leads to the familiar expression~\cite{barci}:
\begin{eqnarray}  
\label{eq:propinv0} 
&&S^{inv}_{(0)}(x-y)={\cal S}_0(x-y)\\
&&\times\exp\left[2ig^2\beta(x-y)-i\frac{g^2}{2}\int_{\gamma(x,y)}\!\!\!\!\!\!\!\!\!\!dz_\mu \int_{\gamma(x,y)}\!\!\!\!\!\!\!\!\!\!dw^\mu \triangle(z-w)+ig^2\gamma^5\int d^2z\int_{\gamma(x,y)}\!\!\!\!\!\!\!\!\!\!dw_\mu (D(z-y)-D(z-x))\epsilon^{\mu\alpha}\partial^z_\alpha\triangle(z-w)\right]\; ,  
\nonumber
\end{eqnarray}
\end{widetext}
In the instanton sectors other than zero, there appear contributions from zero modes of the Dirac operator. We again write the string expression in the source form~(\ref{eq:str}) and consider~(\ref{eq:gener}) for $k=\pm 1$. We will not give here details of this calculation, since it is similar to what was done in~\cite{cadam1} and~\cite{trinst} (see formula (37) of the latter). The obvious difference now is the replacement of the current ${\cal K}^\mu$ defined then as:
\begin{equation}
\label{eq:currk}
{\cal K}^\mu(z;x,y)=g\gamma^\nu\gamma^\mu\partial^z_\nu D(z-x)- g\gamma^\mu\gamma^\nu\partial^z_\nu D(z-y)]\; ,
\end{equation}
and again satisfying:
\begin{equation}
\label{eq:divk}
\partial^z_\mu {\cal K}^\mu(z;x,y)=g(\delta^{(2)}(z-x)-\delta^{(2)}(z-y))\; ,
\end{equation}
with ${\cal K}^\mu-{\cal C}^\mu$. Both currents ${\cal J}^\mu$ and ${\cal K}^\mu$ have the same divergence, so we again obtain $\partial^z_\mu({\cal K}^\mu-{\cal C}^\mu)=0$. An inspection of the exponent of the formula~(37) in~\cite{trinst} shows that all `currents' there become now transverse and the longitudinal --- i.e. gauge dependent --- part of $\triangle_{\mu\nu}$ does not contribute. The evaluation of the resultant expression gives:
\begin{widetext}
\begin{eqnarray}
\label{eq:propinv1}
&&S^{inv}_{(1)}(x-y)=\frac{ig e^{\gamma_E} e^{-i\gamma^5\theta}}{4\pi^{3/2}}\\
&&\times\exp\left[-i\frac{g^2}{2}\int_{\gamma(x,y)}\!\!\!\!\!\!\!\!\!\!dz_\mu \int_{\gamma(x,y)}\!\!\!\!\!\!\!\!\!\!dw^\mu \triangle(z-w)+ig^2\int d^2z\int_{\gamma(x,y)}\!\!\!\!\!\!\!\!\!\!dw_\mu (D(z-y)+D(z-x))\epsilon^{\mu\alpha}\partial^z_\alpha\triangle(z-w)\right]\; .  
\nonumber
\end{eqnarray}
\end{widetext}
To find the infrared asymptotics of $S^{inv}(x,y)$, we have to consider all terms appearing in the exponents of~(\ref{eq:propinv0}) and~(\ref{eq:propinv1}), which are similar, although not identical. Happily, for $\gamma(x,y)$ being a segment of the straight line, the expressions:
\begin{equation}
\label{eq:exp1}
\int d^2z\int_{\gamma(x,y)}\!\!\!\!\!\!\!\!\!\!dw_\mu (D(z-y)\pm D(z-x))\epsilon^{\mu\alpha}\partial^z_\alpha\triangle(z-w)
\end{equation}
turn out to be zero. This may be proved by a direct computation, but can also easily be seen in the following way. If we substitute $w^\mu=(y^\mu-x^\mu)t +x^\mu$ (where parameter $t\in[0,1]$), which means that $dw^\mu=(y^\mu-x^\mu)dt$, and shift the integration variable $z^\mu\rightarrow z^\mu+x^\mu$, the only vector appearing in~(\ref{eq:exp1}), after executing the $z$ integration, is $y^\mu-x^\mu$. This vector has to saturate Lorentz indices of $\epsilon^{\mu\alpha}$, so we obtain a factor $\epsilon^{\mu\alpha}(y_\mu-x_\mu)(y_\alpha-x_\alpha)=0$. 

As regards the first integral in~(\ref{eq:propinv0}) and~(\ref{eq:propinv1}), it may be expressed, for the path in question, by the Meijer's function $G$:
\begin{eqnarray}
\label{eq:mei}
&&\int_{\gamma(x,y)}\!\!\!\!\!\!\!\!\!\!dz_\mu \int_{\gamma(x,y)}\!\!\!\!\!\!\!\!\!\!dw^\mu \triangle(z-w)=\\
&&=\frac{i\pi}{g^2}\left(2+G^{2,1}_{1,3}\left(-\frac{g^2(x-y)^2}{4\pi}\left|\begin{array}{c}\frac{3}{2}\\ 0,1,\frac{1}{2}\end{array}\right.\right)\right)\; ,\nonumber
\end{eqnarray}
and if $(x-y)^2\rightarrow -\infty$, it behaves like $\frac{-i\sqrt{\pi}}{2g}\sqrt{-(x-y)^2}$. This term is responsible for the significant change of the character of the infrared asymptotics: from power decay to exponential decay. This refers to all instanton sectors. Now we are in a position to write down the full form of the IR terms of the propagator $S^{inv}$:
\begin{eqnarray}
S^{inv}_{(0)}(x-y)&\sim& \frac{2e^{-\gamma_E}{\cal S}_0(x-y)}{(-g^2(x-y)^2/\pi)^{1/2}}e^{-\frac{g\sqrt{\pi}}{4}\sqrt{-(x-y)^2}}\; ,\nonumber\\
\label{eq:asinv}\\
S^{inv}_{(1)}(x-y)&\sim& \frac{ig e^{\gamma_E} e^{-i\gamma^5\theta}}{(4\pi)^{3/2}}e^{-\frac{g\sqrt{\pi}}{4}\sqrt{-(x-y)^2}}\; .\nonumber
\end{eqnarray}
As we see, both terms decrease exponentially, with the same euclidean correlation length equal to $\frac{4}{\pi\kappa}$, where $\kappa$ is a mass of the Schwinger boson, but with different power prefactors. In $\text{QED}_3$ such an exponential behavior was suggested to be universal and unavoidable in the infrared, for the gauge invariant propagators defined by the Wilson strings~\cite{mitra}. In the Schwinger Model without instantons, the same result has been obtained in~\cite{barci}. We have now shown, that this phenomenon appears for $k=\pm 1$ too. The asymptotic scale invariance is then broken both by the exponential factor and by the various powers of $-(x-y)^2$ in different topological sectors. One should remember, however, that fermions in the SM are confined, and do not exist as asymptotic particles.

\section{Amplitude for the dressed fermions}
\label{sec:dress}
The alternative form of a gauge invariant propagator is connected with the amplitude for the dressed fermion fields as suggested by Dirac~\cite{dir}: 
\begin{eqnarray}
\tilde{\Psi}(x)&=&e^{-i\int d^2z{\cal F}_\mu(z,x)A^\mu(z)}\Psi(x)\; ,\nonumber\\
\label{eq:dresdir}\\
\tilde{\overline{\Psi}}(x)&=&e^{i\int d^2z{\cal F}_\mu(z,x)A^\mu(z)}\overline{\Psi}(x)\; .\nonumber 
\end{eqnarray}
To ensure the gauge independence of the dressed field, a function ${\cal F}_\mu$ has to satisfy the relation:
\begin{equation}
\label{eq:divF}
\partial^\mu_z{\cal F}_\mu(z,x)=g\delta^{(2)}(z-x)\; .
\end{equation}
For the particular choice:
\begin{equation}
\label{eq:Fstring} 
{\cal F}_\mu(z,x)=g\int_{\gamma(x,\zeta)}\!\!\!\!\!\!\!\!\!\! dw_\mu\delta^{(2)}(z-w)\; ,
\end{equation}
where $\zeta$ is arbitrary (in this case the field $\tilde{\Psi}(x)$ is gauge independent up to the global transformation, because the divergence of ${\cal F}_\mu$ has an extra term canceling, however, for bilinears $\Psi\overline{\Psi}$), one obtains the string version of the invariant propagator~(\ref{eq:sinvp1}). Yet, we are free to consider other choices, as for instance:
\begin{equation}
\label{eq:Flandau} 
{\cal F}_\mu(z,x)=g\partial^z_\mu D(z-x)\; .
\end{equation}
The above form of the dressing is of certain importance, since it is closely related with the problem of path dependence in~(\ref{eq:sinvp1}). If one wished to construct a string, which would simultaneously be path independent and ensuring the gauge invariance, it should be set up of the longitudinal part of the field $A^\mu$ only~\cite{lav}. Such a definition eliminates the $\gamma$ dependence because $A^\mu_L$ constitutes the total derivative of certain scalar function, and the string integral is expressible through the values at its ends $x$ and $y$. In this way we arrive at certain particular choice of ${\cal F}_\mu(z,x)$, for:
\begin{eqnarray}
\label{eq:strlong}
&&\exp\left[-ig\int_{\gamma(x,y)}\!\!\!\!\!\!\!\!\!\! dz_\mu A^\mu_L(z)\right]=\\
&&=\exp\left[-ig\int_{\gamma(x,y)}\!\!\!\!\!\!\!\!\!\! dz_\mu \int d^2w \partial^\mu_z\partial_\nu^z D(z-w)A^\nu(z)\right]=\nonumber\\
&&=\exp\left[ig\int d^2w \partial_\nu^w(D(y-w)-D(x-w))A^\nu(w)\right]
\; ,\nonumber\nonumber
\end{eqnarray}
and this is just, what~(\ref{eq:Flandau}) with~(\ref{eq:dresdir}) mean.\footnote{The choice (\ref{eq:Flandau}) is also minimal in guaranteeing gauge invariance of the fields~(\ref{eq:dresdir}), in this sense, that it is purely longitudinal.}

It should be noted here, that our calculation is fully nonperturbative and fermions are confined. For such a nonperturbative evaluation of the amplitude with dressed fields~(\ref{eq:dresdir}) via Feynman path integral we do not need the existence of asymptotic free fields, and the time nonlocality spoken of in~\cite{lav} is not an obstacle here. 

The simplest way to calculate the propagator with such dressed fields is to replace fermionic source terms in the generating functional~(\ref{eq:gener}) with:
\begin{eqnarray}
&&\overline{\Psi}(x)e^{ig\int d^2z\partial^z_\mu D(z-x)A^\mu(z)}\eta(x)\; ,\nonumber \\
\label{eq:newsource}\\
&&\overline{\eta}(x)e^{-ig\int d^2z\partial^z_\mu D(z-x)A^\mu(z)}\Psi(x)\; .\nonumber
\end{eqnarray}
This allows us to apply directly the formulas from~\cite{trinst} with obvious modifications only. Assume the following form of the gauge field:
\begin{equation}
\label{eq:genpot}
A^\mu(x)=A^{(0)\mu}(x)+\varepsilon^{\mu\nu}\partial_{\nu}b(x)\; ,
\end{equation}
where $A^{(0)\mu}$ is certain potential restricted to the instanton sector $k=0$ and $b$ is the external scalar function, the choice of which is dictated by the
topology of the gauge field. In that way, the whole nonzero winding number of $A^{\mu}$ may be attributed to $\varepsilon^{\mu\nu}\partial_{\nu}b$, $A^{(0)\mu}$
being the trivial topology field. $A^\mu(x)$ standing in the exponents of~(\ref{eq:newsource}) reduces actually to $A^{(0)\mu}(x)$.

Since~(\ref{eq:genpot}) constitutes a simple shift we can now easily pass in~(\ref{eq:gener}) from the functional integration over $A$ to that over $A^{(0)}$. It is known that the coupling term $g\overline{\Psi}\not\!\! A^{(0)}\Psi$ may be gauged away if we introduce new fermion fields defined by the relations:
\begin{eqnarray}
\Psi(x)&=&e^{-ig\not\!\: \partial_x\int d^2z D(x-
z)\gamma^{\mu} A^{(0)}_{\mu}(z)}\Psi'(x)\; ,\nonumber\\ 
\label{eq:gauge1}\\
\overline{\Psi}(x)&=&\overline{\Psi}'(x)e^{ig\gamma^{\mu}\not\!\: 
\partial_x\int d^2z D(x-z) A^{(0)}_{\mu}(z)}\; . \nonumber
\end{eqnarray}
This transformation is an element of $U(1)\otimes U_A(1)$ group, which leaves Lagrangian~(\ref{eq:lagr}) invariant but the same cannot be told about fermionic measure in~(\ref{eq:gener})~\cite{roskies,fuji,bert}. This is a reflection of the anomaly present in the model, leading to the gauge boson mass generation. The other effect of the transformation~(\ref{eq:gauge1}) is the modification of the source terms, which now become:
\begin{eqnarray}
&&\overline{\Psi}(x)e^{ig\gamma^5\int d^2z\partial^z_\mu D(z-x)\epsilon^{\mu\nu}A^{(0)}_\nu(z)}\eta(x)\; , \nonumber\\
\label{eq:sourcemod}\\
&&\overline{\eta}(x)e^{ig\gamma^5\int d^2z\partial^z_\mu D(z-x)\epsilon^{\mu\nu}A^{(0)}_\nu(z)}\Psi(x)\; .\nonumber
\end{eqnarray}
The alteration of the above expressions is the only difference as compared to~\cite{trinst}. This allows us to skip most of the calculations. In the $k=0$ sector the effect of~(\ref{eq:sourcemod}) is to replace ${\cal J}^\mu$ in~(\ref{eq:s0}) with its transverse part:
\begin{eqnarray}
\label{eq:currjt}
{\cal J}^\mu(z;x,y)&\rightarrow &{\cal J}_T^\mu(z;x,y)\\
&=&g\gamma^5\epsilon^{\mu\nu}\partial^z_\nu[D(z-x)- D(z-y)]\; .\nonumber
\end{eqnarray}
The whole procedure becomes then equivalent to taking $\triangle_{\mu\nu}$ in the Landau gauge. It turns out, that the same happens also for nontrivial topological sectors. For $k=\pm 1$, the starting point will be the formula~(37) in~\cite{trinst}. If we recall that both currents ${\cal J}^\mu$ and ${\cal K}^\mu$ have the same divergence, i.e. the same longitudinal parts, we immediately see that the transformations~(\ref{eq:sourcemod}) turn ${\cal K}^\mu$ into ${\cal K}_T^\mu$:
\begin{eqnarray}
\label{eq:currkt}
{\cal K}^\mu(z;x,y)&\rightarrow& {\cal K}_T^\mu(z;x,y)\\
&=&-g\gamma^5\epsilon^{\mu\nu}\partial^z_\nu[D(z-x)+ D(z-y)]\; .\nonumber
\end{eqnarray}
The other term multiplying $\triangle_{\mu\nu}$ in the exponent of the quoted formula~(37) of~\cite{trinst} is already transverse, so again the whole story reduces to the choosing of the Landau gauge. We conclude then, that the whole propagator and, naturally, its infrared asymptotics is the same as found in section~\ref{sec:gvar:subsec:lg} and, after all, the anomalous dimensions found in~(\ref{eq:as0}) and~(\ref{eq:as1}) turn out to have some meaning.

For the choice of ${\cal F}^\mu$ other than~(\ref{eq:Flandau}) one would have to do with the following substitutions:
\begin{eqnarray}
{\cal J}^\mu(z;x,y)&\rightarrow &{\cal J}^\mu(z;x,y)-{\cal F}^\mu(z,x)+{\cal F}^\mu(z,y)
\; ,\nonumber \\
\label{eq:currf}\\
{\cal K}^\mu(z;x,y)&\rightarrow &{\cal K}^\mu(z;x,y)-{\cal F}^\mu(z,x)+{\cal F}^\mu(z,y)
\; .\nonumber
\end{eqnarray}
It should be emphasized, that these new currents are still transverse, although in general other than those given by~(\ref{eq:currjt}) and~(\ref{eq:currkt}), so again only $\triangle^T_{\mu\nu}$ contributes. Thus, the neglecting of $\triangle^L_{\mu\nu}$ is universal for all fields~(\ref{eq:dresdir}), but the transverse contribution is ${\cal F}$ dependent.

\section{Summary and outlook}
\label{sec:sum}
In this section we would like to summarize the obtained results. The solvability of the Schwinger Model allows us to find the infrared behavior and anomalous dimensions for the fermion Green's functions in all instanton sectors. In the case of an ordinary, gauge dependent propagator, the IR anomalous dimension equals to $-1/2$ when we choose the Landau gauge or $\xi/2-1/2$ for the $R_\xi$ gauge. The value of this exponent is independent on the topological sector and may be simply put to zero, if one chooses $\xi=1$.

The gauge independent propagator may be defined in two possible ways: either by the string ansatz adopted from Schwinger or as an amplitude of the dressed fermions suggested by Dirac. In the first case, the correlation function becomes path dependent. For the simplest case of a straight line it may be found explicitly. One observes then, the essential change of the infrared asymptotics: the suppression of the fermion propagator is no longer algebraic but becomes exponential, with the correlation length being proportional to the inverse of the mass of the fermionic bound state --- the Schwinger boson. This phenomenon appears in all instanton sectors, differing only in algebraic prefactors. The $k=0$ contribution decays slightly faster than that for $k=\pm 1$, which seems to be a natural result.

For the gauge independent function defined via Dirac-like dressed fields, the result is dressing dependent. For the dressing chosen in section~\ref{sec:dress}, which is equivalent to constructing the string of the longitudinal part of the gauge field only, we find the propagator to be identical to the gauge variant one calculated in the Landau gauge. This again refers to all instanton sectors.

Undoubtedly, the principal question to be answered is the proper definition of the invariant propagator. In general this definition will probably be not unique, but there should be certain prescriptions for the specific cases. For the `string' propagator the following questions, worth being answered, arise. Firstly, it would be interesting to calculate propagator for paths $\gamma$ other than the segment of the straight line and find whether infrared behavior is sensible to that choice. Secondly, on might think about a systematic study of the possible connection between the fixing of a gauge in the variant function and the choice of a specific path in the invariant one. Some work in this direction was done in~\cite{gaete}. Thirdly, one might ask, whether there are any physical reasons for the choice of the particular path $\gamma$ as more convenient than another.

As higher functions can also be found explicitly in the Schwinger Model in all topological sectors, it would be interesting to construct the two-fermion (i.e. four-point) Green's function in a gauge-invariant way, and investigate both its infrared anomalous dimensions and analytical properties. This analysis my be the more interesting, that this function contains a bound state singularity, the character of which varies between topological sectors.\\

\section*{Acknowledgments}
I would like to thank to Professor J\'ozef Namys{\l}owski for interesting discussions.

\end{document}